\documentclass[journal]{IEEEtran}

\ifCLASSINFOpdf

\else

\fi

\usepackage{bm}
\usepackage{graphicx}
\usepackage{lineno,hyperref}
\usepackage{helvet}
\usepackage{courier}
\usepackage{amsfonts,amssymb}
\usepackage{amsmath}
\usepackage{booktabs}
\usepackage{url}
\usepackage{graphicx}
\usepackage[numbers,sort&compress]{natbib}
\modulolinenumbers[5]
\usepackage{balance}
\begin{document}

\title{Automated Audio Captioning via \\ Fusion of Low- and High- Dimensional Features}

\author{
      \IEEEauthorblockN{
      Jianyuan Sun$^{1,3}$,
      Xubo Liu$^{1}$,
      Xinhao Mei$^{1}$,
      Mark D. Plumbley$^{1}$,\\
      Volkan Kilic$^{2}$,
      Wenwu Wang$^{1}$
     }
     \\

      \IEEEauthorblockN{$^1$Centre for Vision, Speech and Signal Processing (CVSSP), University of Surrey, UK}  \\
      \IEEEauthorblockN{$^2$Department of Electrical and Electronics Engineering, Izmir Katip Celebi University, Turkey}  \\
      \IEEEauthorblockN{$^3$College of Computer Science and Technology, Qingdao University, China}  \\
      }
\markboth{Journal of \LaTeX\ Class Files,~Vol.~14, No.~8, August~2015}%
{Shell \MakeLowercase{\textit{et al.}}: Bare Demo of IEEEtran.cls for IEEE Journals}

\maketitle

\begin{abstract}
Automated audio captioning (AAC) aims to describe the content of an audio clip using simple sentences. Existing AAC methods are developed based on an encoder-decoder architecture that success is attributed to the use of a pre-trained CNN10 called PANNs as the encoder to learn rich audio representations. AAC is a highly challenging task due to its high-dimensional talent space involves audio of various scenarios. Existing methods only use the high-dimensional representation of the PANNs as the input of the decoder. However, the low-dimension representation may retain as much audio information as the high-dimensional representation may be neglected. In addition, although the high-dimensional approach may predict the audio captions by learning from existing audio captions, which lacks robustness and efficiency. To deal with these challenges, a fusion model which integrates low- and high-dimensional features AAC framework is proposed. In this paper, a new encoder-decoder framework is proposed called the Low- and High-Dimensional Feature Fusion (LHDFF) model for AAC. Moreover, in LHDFF, a new PANNs encoder is proposed called Residual PANNs (RPANNs) by fusing the low-dimensional feature from the intermediate convolution layer output and the high-dimensional feature from the final layer output of PANNs. To fully explore the information of the low- and high-dimensional fusion feature and high-dimensional feature respectively, we proposed dual transformer decoder structures to generate the captions in parallel. Especially, a probabilistic fusion approach is proposed that can ensure the overall performance of the system is improved by concentrating on the respective advantages of the two transformer decoders. Experimental results show that LHDFF achieves the best performance on the Clotho and AudioCaps datasets compared with other existing models.
\end{abstract}


\IEEEpeerreviewmaketitle

\section{Introduction}

\IEEEPARstart{A}{utomated} audio captioning (AAC) is a cross-modal translation task that generates a text description for a given audio clip~\cite{DrossosAV17}. AAC has wide application potential. For example, it can help hearing impaired people understand the content of the audio clip. In recent years, the AAC problem has received a lot of attention in the acoustic signal processing and machine learning areas.

Existing AAC models are based on an encoder-decoder architecture~\cite{DrossosAV17, koizumi2020transformer,MeiHLCWWZLKTSPW21}. That is, an audio clip is encoded into a latent embedding representation and aligned with its corresponding natural language description. Then, the latent embedding representation as the input of the decoder to generate the captions. Most existing methods tend to use the pre-processing method on the encoder part to extract the useful information of the latent embedding representation. For example, Tran et al. proposed three learnable processes for encoding, two processes for learning and extracting the local and temporal information, and one to fuse the output of the previous two processes~\cite{tran2020wavetransformer}. Xuenan et al. explored a transfer learning method to learn local and global information~\cite{xu2021investigating}. Where Audio Tagging (AT) and Acoustic Scene Classification (ASC) were employed to represent local and global audio information. Xinhao et al. employed a transformer encoder to access audio information, which pointed out can well capture temporal relationships among audio events~\cite{MeiLHPW21}. Moreover, Chen et al. proposed an interactive audio-text representation method for the audio encoder using contrastive learning~\cite{chen2022interactive}. In addition, some existing methods try to use the transformer decoder or Pre-trained BERT from the natural language processing area as the decoder to generate captions~\cite{Liuprebert22}. Moreover, Xubo et al. proposed a contrastive learning loss to pull together the representations of audio-text paired data in the latent space while pushing apart clusters of unpaired negative data~\cite{LiuHMKTPW21}. AAC is a highly challenging task, in which talent embedding representation involves rich audio information. The talent embedding representations of the encoder output is a typical high-dimension feature representations. It is not difficult to find that existing methods only use the high-dimensional feature representation as the decoder input whatever auxiliary methods are used in the encoder part. However, these high-dimensional approaches may generate the captioning without learned models by learning in existing audio datasets, which lacks robustness and efficiency~\cite{CuiLXZZ13}. In the computer vision area, the low-dimensional approach successfully overcomes the high-dimensional problem of predicting tasks with available training data by learning models, but it only works with specific task types~\cite{CuiLXZZ13}.

In this paper, to solve existing high-dimensional approaches that may predict the audios without learned models by learning in existing audio of different scenes, which lacks robustness and efficiency. We propose a new encoder-decoder framework called the Low and High-Dimensional Feature Fusion (LHDFF) model for audio captioning. In LHDFF, we proposed a new encoder called Residual PANNs (RPANNs) by fusing the low-dimensional feature from the intermediate convolution layer output and the high-dimensional feature from the final layer output of PANNs. To fully learn the information of the low- and high-dimensional fusion feature and high-dimensional feature respectively, we propose dual transformer decoder structures to generate the captions in parallel. Especially, a probabilistic fusion approach is proposed and used to generate the final captions that can ensure the overall performance of the system is improved by concentrating on the respective advantages of the dual transformer decoders.

The remainder of the paper is organized as follows. Our proposed Low and High Dimensional Feature Fusion (LHDFF) model is introduced in Section 2. Section 3 introduces the experiments and results. Finally, section 4 gives the conclusion.

\section{Proposed method}

\begin{figure}[!hbt]\label{fig:MODEL}
  \centering
  \includegraphics[height = 8.73cm, width = 8.85cm]{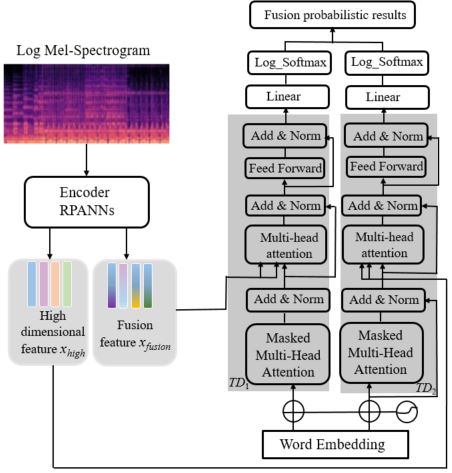}
  \caption{The architecture of the proposed LHDFF. LHDFF consists of an RPANNs encoder and a dual decoder. The RPANNs encoder outputs two embedding features, i.e., the high dimensional feature $x_{high}$ and the fusion feature $x_{fusion}$. Next, $x_{fusion}$ and $x_{high}$ are respectively input to the dual transformer decoder $TD_{1}$ and $TD_{2}$. Each transformer decoder $TD_{1}$ and $TD_{2}$ follows a linear layer and a log softmax layer to output the probability distribution along the vocabulary vector. The final fused probability result is obtained by fusing the log softmax probability distribution of each transformer decoder $TD_{1}$ and $TD_{2}$.}
\end{figure}

In this section, we introduce the architecture of the proposed LHDFF model, which consists of an improved PANNs encoder named residual PANNs (RPANNs) and a dual Transformer decoder. In Fig.~\ref{fig:MODEL}, the overall LHDFF architecture is given.

\subsection{Residual PANNs encoder (RPANNs)}

In LHDFF, an improved PANNs encoder called RPANNs is proposed to extract the audio features. A lot of existing research work has shown that PANNs have strong power in extracting audio signal features for audio pattern recognition tasks such as audio captioning and acoustic scene classification~\cite{sun2022deep}.

The original PANNs involve a CNN10 that consists of four convolutional blocks. Each convolutional block has two convolutional layers with a $3\times3$ kernel size. Moreover, batch normalization and ReLU are used after each convolutional layer. The channel number of each convolutional block is $64, 128, 256$ and $512$. Moreover, an average pooling layer with the kernel size $2\times2$ is applied for downsampling. After the last convolutional block, a global average pooling is applied along the frequency axis to align the dimension of the output with the hidden dimension $D$ of the decoder.

\begin{figure}[!hbt]
  \centering
  \includegraphics[height = 7.73cm, width = 4.65cm]{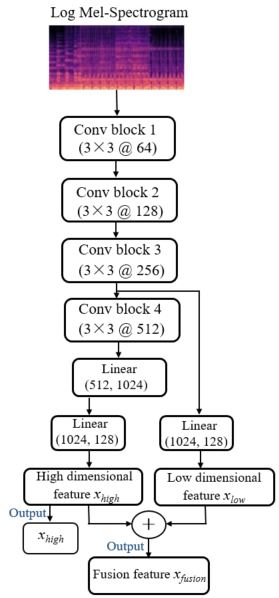}
   \caption{The architecture of the proposed RPANNs. RPANNs consist of four convolutional blocks. The RPANNs encoder outputs two features, i.e., the high dimensional feature $x_{high}$ and the fusion feature $x_{fusion}$.}\label{fig:RPANNS}
\end{figure}

To further improve the performance of PANNs, we propose a new PANNs encoder called RPANNs. The basic architecture of the proposed RPANNs is similar to PANNs. In Fig.~\ref{fig:RPANNS}, we can see the RPANNs consist of four convolutional blocks. The channel number $D$ of each convolutional block is $64, 128, 256$ and $512$. The difference between PANNs and RPANNs is RPANNs fuses the low-dimension feature from the third convolutional block output and the high-dimensional feature from the final layer output. The RPANNs encoder takes the log mel-spectrogram of an audio clip as the input and outputs the high-dimensional feature $I\in \mathbb{R}^{T\times D}$, where $T$ denotes the numbers of time frames, $D$ represents the dimension of the spectral features at each time frame. In our setting of the channel number of each convolutional block, let $x_{3}$ be the output of the third convolutional block, which is a low-dimensional feature. Then, $x_{3}\in\mathbb{R}^{T^{'}\times 256}$, here $D = 256$. Let $x_{final}$ denotes the output of the final layer, which is a high-dimensional feature. The final layer is a linear layer $f_{1024}(\ast)$, then $x_{final}\in\mathbb{R}^{T\times 1024}$, here $D = 1024$. For $x_{3}\in\mathbb{R}^{T^{'}\times 256}$ and $x_{final}\in\mathbb{R}^{T\times 1024}$, $T^{'}\ne T$. In RPANNs, setting the dimensional $D = 128$ of output high-dimensional. Therefore, we can get the low- and high- dimensional fusion feature $x_{fusion}$ and the high-dimensional feature $x_{low}$ as follows.

\begin{equation*}
x_{high} = Relu(f_{128}(x_{final})), x_{high}\in \mathbb{R}^{T\times 128};
\end{equation*}
\begin{equation*}
x_{low} = Relu(f_{128}(x_{3})), x_{low}\in \mathbb{R}^{T^{'}\times 128};
\end{equation*}
\begin{equation*}
x_{fusion} = x_{high} \oplus x_{low}, x_{fusion}\in \mathbb{R}^{T\times 128}.
\end{equation*}
Where $T^{'}\ne T$, the method of patch $0$ is used to make the dimension of $T^{'}$ and $T$ the same. Moreover, in the encoder, we take $x_{fusion}$ and $x_{high}$ as the input of the dual decoder, respectively.

\subsection{Dual decoder}

To fully explore the information of the low- and high-dimensional fusion feature and high-dimensional feature respectively, the dual decoder is proposed and used. Moreover, $x_{fusion}$ and $x_{high}$ as the input of dual decoder, respectively. The dual decoder consists of eight parts, a word embedding layer, two standard Transformer decoders, two linear layers, two softmax probability layers, and a fusion softmax probability layer, which is shown in Fig.~\ref{fig:MODEL}. The word embedding layer can be seen as a word embedding matrix $W$ with size $V \times d$, where $V$ denotes the size of the vocabulary and $d$ represents the dimension of each word vector. During training, the word embedding layer is randomly initialized and kept frozen.

The architecture of the dual decoder can be seen in Fig.~\ref{fig:MODEL}, which includes two standard transformer decoders. Transformer decoder is a state-of-the-art model in the areas of natural language processing (NLP)~\cite{VaswaniSPUJGKP17}, which has shown the effectiveness of the AAC~\cite{MeiHLCWWZLKTSPW21, MeiLHPW21}. Here, we use the notation $TD_{1}$ and $TD_{2}$ to denote two transformer decoders involved in the dual decoder, respectively. Where, the word embedding matrix $W$ together with the fused feature $x_{fusion}$ as the input of $TD_{1}$. Moreover, the word embedding matrix $W$ together with the high dimensional feature $x_{high}$ as the input of $TD_{2}$. As shown in Fig.~\ref{fig:MODEL}, each transformer decoder $TD_{1}$ and $TD_{2}$ includes a multi-head attention mechanism. Empirically, we set the number of head attention is four for each transformer decoder, and the dimension of the hidden layer is $128$. Each transformer decoder $TD_{1}$ and $TD_{2}$ follows a linear layer and a softmax probability layer to output the probability distribution along the vocabulary vector.

Let $m$ represent the number of words in the dictionary for a dataset. $f_{TD_{1}}$ and $f_{TD_{2}}$ denote the linear layer following the Transformer decoder $TD_{1}$ and $TD_{2}$. We can get the output of each linear layer for each Transformer decoder $TD_{1}$ and $TD_{2}$, as follows:
\begin{equation*}
x_{TD_{1}} = f_{TD_{1}}(x_{fusion}), x_{TD_{1}}\in \mathbb{R}^{T \times m};
\end{equation*}
\begin{equation*}
x_{TD_{2}} = f_{TD_{2}}(x_{high}), x_{TD_{2}}\in \mathbb{R}^{T \times m}.
\end{equation*}

Then, we can get the final fused probability distribution $P_{fusion}$ for the vocabulary by fusing the probability distribution of each transformer decoder $P_{TD_{1}}$ and $P_{TD_{2}}$, as follows:
\begin{equation*}
P_{TD_{1}} = logsoftmax(x_{TD_{1}}), P_{TD_{1}}\in \mathbb{R}^{T\times m};
\end{equation*}
\begin{equation*}
P_{TD_{2}} = logsoftmax(x_{TD_{2}}), P_{TD_{2}}\in \mathbb{R}^{T\times m};
\end{equation*}
\begin{equation*}
P_{fusion} = P_{TD_{1}} \oplus P_{TD_{2}}, P_{fusion}\in \mathbb{R}^{T\times m}.
\end{equation*}

The training objective of RPANNs is to optimize the cross-entropy (CE) loss defined in terms of the model parameters $\theta$ as:
\begin{equation*}
L_{CE}(\theta) = -\frac{1}{T}\Sigma_{t=1}^{T} logP_{fusion}(y_{t}|y_{1:t-1}, \theta)
\end{equation*}
where $y_{t}$ represents the ground truth word at time step $t$.
\section{Experiments}

To verify the performance of the proposed LHDFF in AAC, some comparative and validation experiments were designed and performed on Clotho~\cite{drossos2020clotho} and AudioCaps~\cite{kim2019audiocaps} datasets.

\subsection{Datasets}
\subsubsection{Clotho and AudioCaps}

Clotho~\cite{drossos2020clotho} is a classical audio captioning dataset that audio clips are from the Freesound archive. Each audio clip has five captions annotated by different Amazon Mechanical Turk persons. The duration of all the audio clips ranges from $15$ to $30$ seconds. All the audio captions contain $8$ to $20$ words. In our experiment, the Clotho v2 is used that releases for the Task $6$ of DCASE $2021$ Challenge, which contains $3839$ development, $1045$ validation, and $1045$ evaluation split. To comply with the settings of the majority of existing methods, we also merge the development and validation split together, which refers to the training dataset with $4884$ audio clips. The evaluation split is selected as the test dataset with $1045$ audio clips. Moreover, each audio clip combines one of its five captions as a training sample in the training dataset.
%
\begin{table*}[tbp]
   \centering
  \caption{Ten comparative experimental results on Clotho and AudioCaps datasets. Where the Baseline method does not use reinforcement learning (RL).}
  \begin{tabular}[]{c| c| c c c c c c c c c}
  \hline
  \textbf{Dataset}     &\textbf{Model}  &BLEU1  &BLEU2  &BLEU3  &BLEU4  &ROUGEL  &METERO  &CIDEr &SPICE  &SPIDEr   \\ \hline
                        & Baseline~\cite{MeiHLCWWZLKTSPW21}~(without RL) &0.561  &0.364  &0.243 &\textbf{0.159}    &0.375    &0.172  &0.391  &0.120    &0.256  \\
       Clotho           & CL4AC model~\cite{LiuHMKTPW21}   &0.553   &0.349    &0.226    &0.143    &0.374    &0.168   &0.368    &0.115    &0.242   \\
                        & AT-CNN10~\cite{xu2021investigating} &0.556   &0.363    &0.242    &0.159    &0.368    &0.169   &0.377    &0.115    &0.246   \\  \hline
                        &LHDFF~(Our model only fusi-fea) &0.570   &0.370   &0.246  &0.158  &0.378  &0.174  &0.401  &0.120  &0.261 \\
                        & LHDFF~(Our model) &\textbf{0.570} &\textbf{0.370} &\textbf{0.247} &\textbf{0.159}  &\textbf{0.378}  &\textbf{0.175} &\textbf{0.408} &\textbf{0.122} &\textbf{0.265} \\ \hline
                        & Baseline~\cite{MeiHLCWWZLKTSPW21}~(without RL) &0.667     &0.491     &0.350   &0.248    &0.468    &0.229  &0.643  &0.165 &0.404 \\
      AudioCaps         & Pre-Bert~\cite{Liuprebert22}  &0.667    &0.491    & 0.354    & 0.247    & 0.475   & \textbf{0.232}   & 0.654    & 0.167 & 0.410 \\
                        & AT-CNN10~\cite{xu2021investigating} &0.655   &0.476   &0.335    &0.231    &0.467  &0.229  &0.660  &0.168  &0.414  \\    \hline
                        & LHDFF~(Our model only fusi-fea)  &\textbf{0.674}     &0.500     &0.367   &0.263  &0.481 &\textbf{0.231}  &0.666 &\textbf{0.171}    &0.419   \\
                        & LHDFF~(Our model) &\textbf{0.674} &\textbf{0.502} &\textbf{0.368} &\textbf{0.267} &\textbf{0.483} &\textbf{0.232} &\textbf{0.680} &\textbf{0.171} &\textbf{0.426}    \\
                        \hline
  \end{tabular}
  \label{tab:audioresult}
\end{table*}
AudioCaps~\cite{kim2019audiocaps} is the largest audio captioning dataset, which includes $50k$ audio clips with a duration of $10$ seconds. AudioCaps is partitioned into three parts with $49274$ audio clips for the training dataset, $497$, and $957$ audio clips for the validation and test dataset, respectively. Each audio clip has one caption in the training dataset, and each audio clip involves $5$ captions in the validation and test datasets, respectively. The length of the captions ranges from $3$ words to $20$ words.

\subsection{Data pre-processing}

For the audio clips, we use a 1024-point Hanning window with a hop size of 512-points to obtain 64-dimensional log mel-spectrograms as the LHDFF input features. Moreover, the SpecAugment~\cite{park2019specaugment} method is employed to augment the training data, which augments the log mel-spectrogram of an audio clip by using the zero-value masking and mini-batch based mixture masking~\cite{wang2021specaugment++}. For all the captions in the Clotho and AduioCapts datasets, the captions are transformed to lowercase and removed punctuation. Moreover, we pad two special tokens <sos> and <eos> at the beginning and end of each caption.

\subsection{Experimental setups}
The proposed LHDFF model is trained using the Adam optimizer~\cite{kingma2014adam} with a batch size of $32$. We set the model training epoch to $30$ with an initial learning rate (lr) of $5 \times 10^{-4}$, because the model performs best on the validation at epoch 30. In the first $5$ epochs, warm-up is applied to increase the initial lr linearly. Then, the lr is decreased to $1/10$ every $10$ epochs. For all the captioning, the Word2Vec model is used to pre-trained word embedding in Clotho and AudioCaps~\cite{mikolov2013efficient}.

\subsection{Metric method}

Existing mainstream metric methods for evaluating the performance of the audio captioning models including machine translation metrics: BLEUn~\cite{papineni2002bleu}, METEO~\cite{agarwal2007meteor}, ROUGEL~\cite{rouge2004package} and generated captioning metrics: CIDEr~\cite{vedantam2015cider}, SPICE~\cite{anderson2016spice}, SPIDEr~\cite{liu2017improved}. BLEUn mainly measures the n-gram precision of a generated text. METEOR is a word-to-word matching-based harmonic mean of recall and precision. Based on the longest common subsequence, ROUGEL computes F-measures. The term frequency-inverse document frequency (TF-IDF) of the n-gram is taken into account by CIDEr. SPICE takes captions from scene graphs and uses them to determine F-score. SPICE takes captions from scene graphs and uses them to determine F-score. SPIDEr denotes the mean score of CIDEr and SPICE.

\subsection{Results}
\subsubsection{Comparison with baseline methods}
To effectively verify the validity of the proposed LHDFF model, for the Clotho dataset, we compared LHDFF with three representative baseline methods, i.e., a baseline model is a classical encoder-decoder based audio captioning system~\cite{MeiHLCWWZLKTSPW21}, CL4AC model~\cite{LiuHMKTPW21}, and AT-CNN10~\cite{xu2021investigating}. Where, the baseline model~\cite{MeiHLCWWZLKTSPW21} contains a PANNs encoder and a transformer decoder, which uses reinforcement learning techniques and obtains 3rd place in DCASE 2021 Task6. For fairness in comparison, we only report results of the baseline model that does not use reinforcement learning. CL4AC~\cite{LiuHMKTPW21} is proposed based on contrasting learning to reduce the domain difference by learning the correspondence between the audio clips and captions. AT-CNN10~\cite{xu2021investigating} method mainly uses transfer learning to initial the parameters of the encoder of the audio captioning by learning the local feature from the Audio Tagging task and the global feature from the Acoustic Scene Classification. For the AudioCaps, in addition to the baseline model~\cite{MeiHLCWWZLKTSPW21} and AT-CNN10~\cite{xu2021investigating}, we compared a Pre-Bert method for audio captioning that uses the Pre-trained BERT language as the decoder~\cite{Liuprebert22}. Table~\ref{tab:audioresult} shows the performance of all comparison methods and the proposed LHDFF model. It is easy to find that the proposed LHDFF model achieves the best performance compared with the other models.

\textbf{High- vs. Fusion embedding feature} To verify the validity of the low-high dimensional fusion embedding features, as shown in Table~\ref{tab:audioresult}, we report the result of the proposed LHDFF model that only uses and inputs the fusion embedding feature to the transformer decoder, that is see the result of the LHDFF (only fusi-fea) in Table~\ref{tab:audioresult} on the Clotho dataset. Moreover, we report the result of the model that only used and inputs the high-dimensional to the transformer decoder, that is see the results of the Baseline in Table~\ref{tab:audioresult} on the Clotho dataset. Because the difference between the baseline and LHDFF model is the baseline model only uses the high-dimensional feature as the input of a single transformer decoder, the LHDFF model used the high-dimensional and low-high dimensional fusion features as the input of the dual transformer decoder respectively. From the result, it is easy to find that the performance of the baseline model only uses the high-dimensional feature worse than the LHDFF model which only uses low-high dimensional fusion features. These results show that the model that combines the low- and high-dimensional features can learn more useful in formations, which can improve the robustness and efficiency of the model that only uses the high-dimensional feature. Then, higher prediction accuracy can be obtained.

\textbf{Dual decoder vs. Single decoder} To verify the efficiency of the dual decoder, we reported the result of the proposed LHDFF that uses a dual transformer decoder to output the final probability result, where the high-dimensional and low-high dimensional fusion features as the input of the dual transformer decoder respectively. Moreover, we are given the results of the baseline and the LHDFF method that are only use a single decoder to output the final results by using the high-dimensional feature and fusion-dimensional feature, respectively. From the comparison results between the baseline and LHDFF~(only fusi-fea), we can find the model that uses the lower-high dimensional fusion feature and inputs a single decoder to outperform the model uses the high dimensional feature and inputs a single decoder. More importantly, the LHDFF used the dual decoder outperforms the LHDFF~(only fusi-fea) only used a single decoder. The idea of proposing a dual decoder is similar to ensemble learning, that is, using multiple learning algorithms to get better performance than could be obtained from any of the constituent learning algorithms alone.

\section{Conclusion}

Few existing AAC models have investigated the performance of combining low- and high-dimensional audio features, as we know, the low-dimensional approach successfully overcomes the high-dimensional problem of generating the audio captions with available training data by learning encoder-decoder models, but it only works with specific audio scene types. On the other hand, although the high-dimensional approach may predict the audio captions by learning from existing audio captions, which lacks robustness and efficiency. To deal with these challenges, a fusion formulation that integrates low- and high-dimensional features AAC model is proposed called LHDFF. Experimental results show that LHDFF achieves the best performance on the Clotho and AudioCaps datasets compared with other existing models.

\section*{Acknowledgement}
This work is partly supported by a Newton Institutional Links Award from the British Council and the Scientific and Technological Research Council of Turkey (TUBITAK), titled Automated Captioning of Image and Audio for Visually and Hearing Impaired (Grant numbers 623805725 and 120N995), a grant EP/T019751/1 from the Engineering and Physical Sciences Research Council (EPSRC).

\bibliographystyle{IEEEtran}
\bibliography{mybibfile}

\end{document}